\documentstyle[aasms,12pt,epsf]{article}

\def \aa    #1 #2   {{ A\&A\/}, {#1}, {#2}}
\def \aas   #1 #2   {{ A\&AS\/}, {#1}, {#2}}
\def \aj    #1 #2   {{ AJ\/}, {#1}, {#2}}
\def \apj   #1 #2   {{ ApJ\/}, {#1}, {#2}}
\def \apjl  #1 #2   {{ ApJ Lett.\/}, {#1}, {#2}}
\def \apjs  #1 #2   {{ ApJ Suppl.\/}, {#1}, {#2}}
\def \mnras #1 #2   {{ MNRAS\/}, {#1}, {#2}}
\def \prl   #1 #2   {{ Phys. Rev. Lett.\/}, {#1}, {#2}}
\def \nat   #1 #2   {{ Nature\/}, {#1}, {#2}}
\def \com   #1 #2   {{ Comments Astrophys.\/}, {#1}, {#2}}
\def \sast  #1 #2   {{ Soviet Astron.\/} {#1}, {#2}}
\def \sastl #1 #2   {{ Soviet Astron. Lett.\/}, {#1}, {#2}}
\def \astr  #1 #2   {{ Astrophysics\/}, {#1}, {#2}}

\begin{document}

\baselineskip 0.8cm

\title{PERCOLATION ANALYSIS OF A WIENER RECONSTRUCTION 
	OF THE IRAS 1.2 Jy REDSHIFT CATALOG}
\author{Capp Yess and Sergei F. Shandarin \\
	Department of Physics and Astronomy\\
        University of Kansas, Lawrence, KS 66045\\
	\centering and\\
	Karl B. Fisher\\
	Institute for Advanced Studies\\
	Olden Lane, Natural Sciences, Bldg E,\\
	Princeton, NJ 08540}



\begin{abstract}

We present percolation analyses of Wiener Reconstructions of the IRAS 
1.2 Jy Redshift Survey. There are ten reconstructions of galaxy density 
fields in real space spanning the 
range $\beta= 0.1$ to $1.0$, where ${\beta}={\Omega^{0.6}}/b$, $\Omega$ is the 
present dimensionless density and $b$ is the bias factor. Our method 
uses the growth of the largest cluster statistic to characterize the topology 
of a density field, where
Gaussian randomized versions of the reconstructions are used as standards 
for analysis. For the reconstruction volume of radius, $R {\approx} 100 
h^{-1} $ Mpc, percolation analysis reveals a slight `meatball' topology for the 
real space, galaxy distribution of the IRAS survey.

{\it Subject headings}: cosmology-galaxies:clustering-methods:numerical

\end{abstract}

\newpage

\section{Introduction}

Quantifying the distribution of galaxies in the visible universe has been 
one of the primary objectives of the study of the Large Scale Structure of the 
universe for several decades. With the compilation of early two dimensional 
galaxy catalogs (for example, Zwicky {\it et al.} (1961) and Shane \& Wirtanen 
(1967)), astronomers noted structures indicative of clustering and evolution. 
More recent redshift surveys (see, Davis {\it et al.} (1982); de Lapparent, 
Geller \& Huchra (1986); Giovanelli \& Hayes (1985); Tully \& Fisher (1987); 
Lawrence {\it et al.} (1996)) 
produced three-dimensional galaxy distributions and revealed structures such 
as voids (for example, Kirshner {\it et al.} (1981)), and filaments and sheets 
of galaxies (\cite{Gel-Huc89}). Angular and spatial correlation functions 
(\cite{Pee80} and references therein ) 
were employed as initial attempts to distinguish the galaxy 
surveys from random distributions and from comparisons with theoretical models. 

In 1982, Zel'dovich proposed that the large scale structure of the universe 
could be characterized by its topology, and a multitude of statistical 
measures have been developed since then to quantify the topology of a 
distribution: the percolation threshold (\cite{Sh83}; \cite{Sh-Z83}); the genus 
(\cite{Got-Mel-Dic86}); the contour crossing (\cite{Ryd88}); random walk 
statistics (\cite{Bau93}); Minkowski functionals (\cite{Mec-Buc-Wag94}); and 
minimal spanning tree characteristics (\cite{Bha-Spl96}). 
Many obstacles have been overcome in the refinement of the measures above 
including boundary and selection effects, discreteness, local biasing, and 
error estimation. For example, recent topological analyses of the CfA 
redshift survey by percolation (\cite{deL-Gel-Huc91}) and genus 
(\cite{Vog-etal94}) report important and complementary findings. 
The topology and power spectrum of the IRAS survey has been 
examined in the context of the Queen Mary and Westfield College, Durham, 
Oxford and Toronto (QDOT) survey by Moore {\it et al.} (1992). 
The results of every report cited above are consistent with the findings of 
Gott {\it et al.} (1989): On scales significantly larger than the correlation 
length the topology is sponge-like. A sponge topology is characterized 
by equivalent, over- and under-dense, multiply connected regions both of which 
percolate and are completely interlocked.  
These findings are consistent with the scenario that large scale structure 
developed from initial random Gaussian density fluctuations present at the 
epoch of recombination. In addition, for smoothing lengths 
comparable to or smaller than the correlation length slight shifts towards a 
`meatball' 
(\cite{Got-etal89}; \cite{Moo-etal92}) and `bubble' (\cite{Vog-etal94}) 
topology have been reported.  The scope of the upcoming Sloan Digital Sky 
Survey (\cite{Gun-Wei95}) promises even more significant results.

We use a percolation analysis that tracks two parameters in order to measure 
the connectivity of a Wiener reconstruction of the IRAS 1.2 Jy Redshift 
Survey and to estimate the spectral index of its associated power spectrum. 
The Wiener filter (WF) takes a three dimensional, redshift map with the 
excluded zones filled by interpolation and converts it to a noise-free, full 
sky, real space map.
Our percolation code computes the normalized volume of the largest structure 
as a function of the filling factor for a topological comparison. The 
number of clusters statistic has been shown to be sensitive to the index of 
the power spectrum for a simple power law relationship (\cite{Yes-Sh96}). 
The method has been used for studying the properties of voids 
(\cite{Sah-Sat-Sh94}) and the geometry of mass clumps (\cite{Sat-Sah-Sh96}) in 
cosmological N-body simulations.  The 
largest structure can be an over-dense region (cluster \footnote {The terms 
cluster and void in this context refer to high and low density regions 
respectively and not to the common astronomical meanings. Also, for the IRAS 
survey the terms under-dense and over-dense refer to galaxy densities; 
whereas, for N-body simulations the terms refer to mass densities.}) or an 
under-dense region (void). By comparing the growth of the largest structures 
(as functions of the filling factor) in a distribution with their growth in a 
randomized version of the distribution the distribution topology can be 
characterized. A Gaussian randomized version of a distribution is by definition 
a structureless field with the same power spectrum as its parent, so that any 
distribution that percolates at a lower filling factor than its randomized 
version is considered more connected than a random field, and hence, 
a connected network; whereas, any field that percolates at a higher filling 
factor than its randomized version is considered isolated or clumpy.  In 
addition, reconstructions of N-body simulations (with and without Wiener 
filters) over the spectral range $n= -2, -1$ and $0$ are analyzed for 
comparison. 

In \S2 we describe the the density fields derived from Wiener
reconstructions of the IRAS 1.2 Jy 
Redshift Survey (\cite{Fis-etal95a}) and randomized versions 
which preserve the underlying power spectrum but are Gaussian fields. 
In \S3 we detail the percolation method and parameters we have 
developed for analysis and comparison. Our results are presented and 
scrutinized in \S4 with conclusions to follow in \S5.

\section{The Density Fields}

The core of this study are Wiener reconstructions of the {\it real} space 
density field formed from the IRAS 1.2 Jy Survey
(\cite{Fis-etal95b}). This survey is an extension of the 1.936 
Jy flux limited survey of Strauss {\it et al.} (1992) compiled from 
the Infrared Astronomical Satellite Point Source Catalog (1988; PSC). The 
survey contains 5321 galaxies which cover 87.6 per cent of the sky. The 12.4 
per cent of the 
sky that is missing from the survey is the Zone of Avoidance ( $\pm 5\deg$  
from the galactic plane) and a few confused regions or areas lacking coverage. 
For specifics of the galaxy selection criteria, the participating telescopes, 
data reduction techniques and results, the derived selection function, and 
the galaxy distribution see Fisher {\it et al.} (1995b).

\subsection{Wiener Reconstructions}

A Wiener reconstruction method is used to convert an interpolated redshift 
density field to a real space density field while suppressing noise. This 
approach is valid in the context of linear theory implying significant 
smoothing which is an aspect of the filter.
The Wiener filter reconstruction method has been employed in 
many fields (\cite {Ryb-Pre92}) to enhance a signal in the presence of noise. 
In cosmology, one use of the method is the minimum variance 
reconstruction of the real space 
density function from an incomplete and sparsely sampled 
galaxy distribution in redshift space. The process depends upon 
an expectation of the clustering properties of the real underlying density 
field being probed by the galaxy survey. In the cosmological case, the 
underlying density field is assumed to be Gaussian up to near the stage 
of non-linearity. 

The WF algorithm as it applies to the IRAS 1.2 Jy galaxy survey
is discussed in detail in Fisher {\it et al.} 1995a; for completeness
we give a brief summary here.  The filtering is applied to a three dimensional 
decomposition of the redshift space galaxy density field in an orthogonal 
basis set of the spherical harmonics and spherical Bessel functions. The 
decomposition is truncated with 
$l$ ranging from $0\le l \le 15$ (with $-l \le m \le l$) and 
$0\le k_nr \le 100$ as a compromise between resolution and the
number of expansion coefficients.
The WF reconstruction technique depends on the assumed
linear theory growth parameter $\beta=\Omega^{0.6}/b$ where
$\Omega$ is the the current cosmic density and $b$ is the
linear bias parameter. We investigate a set of
ten reconstructions spanning the range $\beta=0.1$ to $\beta=1.0$. 
In each case, the real space density field is reconstructed on a $64^3$ grid 
with sides of length $200 h^{-1} $ Mpc (20,000 km/s).  

To compute the Wiener reconstructions the first step is to compute the 
redshift space harmonics in the spherical harmonics and spherical Bessel 
functions basis. This is analogous to computing the Fourier components in the 
analysis of the power spectrum. 
The redshift harmonics are distorted from the values that would
be measured in a perfect real space galaxy distribution. First,
the actual galaxy distribution is sparsely sampled and this
results in a statistical uncertainty or shot noise in the
estimated harmonics.  Second, peculiar velocities 
introduce a systematic distortion due to the coherent 
infall and outflow around over-dense and under-dense regions. 
In linear theory, this redshift distortion
can be computed if the value of $\beta$ is known; the
spherical basis function is convenient here since the
distortion is in the form of a matrix which couples the
radial modes of the expansion.

In the absence of shot noise, the real space harmonics could
be recovered by a direct inversion of the coupling matrix.
Shot noise makes this inversion highly unstable. The Wiener
filter is a smoothing algorithm which is designed to
make the inversion in the presence of noise optimal in the
sense of minimum variance. It depends on the `prior' which
is the knowledge of the clustering of the underlying field.
Essentially, the Wiener filter is the ratio of the variance
in the signal (determined from the assumed prior power spectrum)
to the sum of the variance in both the signal and noise (determined
by the amplitude of the shot noise). 

The result is a density field in real space centered on the local group. Two
important properties of the reconstruction method are an effective smoothing 
of the resultant field which increases with radius due to the limited 
resolution caused by truncating the harmonic expansion at $l_{max}= 15$, and 
the increased 
attenuation of the signal as a function of radius because the Wiener filter is 
dependent on the shot noise. The direct implication of truncating the harmonic 
expansion at $l_{max}$ for fields sampled on a $64^3$ grid is that at 
distances of $R= 30$ mesh units ($\approx 100 h^{-1} $Mpc) the minimum 
resolution is approximately $R\times (\pi/l_{max})\approx 6$ mu The effects 
of smoothing will be examined in detail in section \S4.  For detailed 
explanations of the Wiener filter reconstruction method for different response 
functions see Fisher {\it et el.} (1995a) and Zaroubi {\it et al.} (1995).

In addition to the reconstructions of the IRAS data, we produced 
reconstructions of cubic ($L^3$) density fields derived from N-body 
simulations with power law initial spectrum ($P(k)=Ak^n$) for $n= -2, -1,$ and 
$0$ evolved to the stage where scales of the size L/4 were approaching 
nonlinearity (\cite{Mel-Sh93}). 
Assuming that the rms fluctuation in the number of galaxies is approximately 
equal to the rms mass fluctuation, both are unity within spheres of radius 
$8 h^{-1} $ Mpc.  So, by identifying the stage where L/4 becomes nonlinear with 
the present, a rough estimate of the size of a mesh cell is $3.1 h^{-1} $ Mpc.  
These reconstructions were produced with and 
without Wiener filtering in order to systematically study the effects of 
harmonic expansion and Wiener filtering. 
In addition, the N-body simulations are reconstructed with the IRAS WF 
and not a WF based on the clustering and noise in the simulations.  
Using all the particles from the simulations to compute their 
harmonics and smoothing with a WF which corresponds to the sampling density of 
the IRAS 1.2 Jy survey assures that the N-body reconstructions show the 
same resolution as the IRAS reconstructions.  
The volume of the N-body reconstructions was chosen to match the local mass 
density of the IRAS galaxy distribution at 500 km/s ($5 h^{-1}$ Mpc), 
approximately $0.045/h^{-3} $ Mpc$^3$ to allow for visual comparison.

\subsection{Gaussian Randomizations}

We produce Gaussian random fields by two methods in this study. 
Reconstructions are expanded in a spherical harmonic basis set and are also 
randomized in this basis; whereas, original N-body simulation density fields 
which are not reconstructed are 
Fourier transformed and then randomized in k-space. The crux of
both processes is the randomization of the phases while retaining the original 
power spectrum of the parent field. For reconstructions (with and without 
Wiener filtering) this is accomplished by multiplying the density, 
$\rho_{lmn}$, for $m>0$ by $e^{i\phi}$, where $\phi$ is a random variable in 
the range $0 \leq \phi \leq 2\pi$. The $m=0$ term is multiplied by 
$\sqrt{2} \cos \phi$, and the $m<0$ terms are determined using a reality 
condition for $\rho$. The randomized versions of fields derived from N-body 
simulations are created by multiplying the components 
of all k-space vectors by $\cos \phi$, and a reality condition again assigns 
values to coefficients in the lower half of k-space.

\section{Percolation Method}

The percolation methodology \footnote{For a detailed description
and evaluation of the percolation technique used in this study see 
Yess \& Shandarin (1996).} we employ analyzes galaxy (IRAS) or mass (N-body) 
distributions as well as void distributions. The intent is to characterize the 
topology of both distributions, and to estimate the slope of the 
power spectrum of the density field.  The discriminator between mass 
sites and voids is the density threshold, and it is smoothly varied 
to establish contours separating clusters and voids. Void and galaxy 
percolation are analogous to mass percolation so for simplicity percolation 
will be discussed in the context of mass percolation except where distinction 
is needed. As the density threshold is varied three parameters are tracked: 
the filling factor, the volume of the largest structure (for both over-dense 
and under-dense structures), and the number of isolated structures.   

The filling factor is the fractional volume of all mass sites identified 
in the distribution for a given density threshold. It is equivalent to 
the cumulative distribution function for the clusters and the volume fraction 
of Gott {\it et al.} (1989) for Gaussian distributions.  For clusters, the 
filling factor grows from a minimum value of zero to a maximum of one as 
the density threshold is systematically lowered.  The filling factor 
serves as the independent variable for our functions to allow for a fair 
comparison between different density fields.  

The second parameter, the volume of the largest cluster, is a stable 
indicator of the percolation transition and is used to assess the 
topology of the field. The volume is reported in units of the filling factor 
(the ratio of the largest cluster to the total volume of all clusters)
as a function of the filling factor, and a rapid increase in the 
volume indicates the filling factor associated with the percolation 
transition. This transition represents a change from a clumpy to a 
connected topology for the field.  For clusters it is a change from a 
meatball to a sponge topology, and for voids it is a change from a 
bubble to a sponge topology. Gaussian fields are used as standards of 
comparison to characterize the topology of density fields. A field which 
percolates at a smaller filling factor than its Gaussian counterpart 
displays a shift towards a connected topology, while a field that 
percolates at higher filling factors displays a shift towards a clumpy 
topology. 

The number of clusters statistic is sensitive to the slope of the power 
spectrum of a field (\cite{Yes-Sh96}). This implies that for a field described 
by a power 
spectrum of the form $P(k)= Ak^n$ that the number of clusters statistic 
is sensitive to the spectral index, $n$. In fact, the maximum of the 
statistic is a function of $n$, and can be used to estimate the effective 
spectral index of a mass distribution.

\section{Results}

Topological analysis of modern redshift surveys has focused on the two 
aspects of the galaxy distributions mentioned above: a quantitative assessment 
of the connectedness of the structure, and the slope of the power spectrum. 
In addition to the 
difficulties of assessing boundary effects and error estimations, a major 
obstacle for all current methods employed to describe the spatial distribution 
of galaxies is the lack of resolution resulting from the sparse sampling 
achieved in existing surveys. 

The resolution of any representation of a galaxy survey is ultimately a 
function of the mean galaxy density of the survey and the chosen smoothing 
method. In this respect the IRAS 1.2 Jy survey presents good prospects 
with the average galaxy number density higher 
than the QDOT survey value in the region $R\approx 100 h^{-1} $ Mpc. 
However, different groups have utilized different smoothing 
routines to produce density fields from the galaxy distributions of the 
surveys. For example, the smoothing methods of Moore {\it et al.} (1992) 
in their analysis of the QDOT survey are typical, but differ significantly 
from those utilized in the spherical harmonic reconstruction of our data. 
Moore and collaborators used a constant Gaussian filtering width determined 
by the inter-galactic spacing at the edge of the QDOT survey, $\lambda= 
[S(r_{max})]^{-1/3}$, where $S(r)$ is the radial selection function. In a 
magnitude limited sample this choice ensures that 
the density field is not under-sampled while providing an unprecedented number 
of resolution elements for the QDOT survey. In contrast, the spherical 
harmonic reconstruction of the IRAS 1.2 Jy survey implies a variable 
smoothing with radius due to the finite cutoff of $l$ in the spherical 
harmonic expansion. In addition, the Wiener filter suppressed 
the amplitude of the field as a 
function radius  to mitigate the effects of
increasing shot noise (as determined by the selection function). 
The effect of the variable smoothing in the density 
field is evident in the results reported below. 

Like all statistical measures our parameters are sensitive to the resolution 
and number density of the data, but they are relatively robust with respect 
to boundary effects and scale. 
Since the volume of the largest structure is normalized to the filling factor, 
and the number of clusters statistic can be normalized to the volume of the 
survey,  the geometry and size of the survey does not determine their 
analytic behavior. We exploit the stability of our parameters by 
analyzing spherical subregions of the survey in order to examine the 
effects of variable smoothing and any local bias against a fair sample. 
A measure of the stability of our parameters are the errors presented for the 
results from N-body simulations with multiple 
realizations. In all instances the error bars represent $1\sigma$ 
deviations over four realizations.  We do not estimate errors in the 
results for the IRAS reconstructions, but rather rely on trends in the versions 
varying in $\beta$, over the range $0.1 \le \beta \le 1.0$, to determine 
conclusions.

\subsection{Largest Structures}

Largest structure results for all versions of the Wiener Reconstructions are 
shown in Figure 1 for both clusters and voids. The top panels show the growth 
of the largest structures for a field sampled on a cubic grid, 64 mu 
($200 h^{-1} $ Mpc) to a side \footnote{ The result of a spherical harmonic 
reconstruction is a spherical field, and the ($64^3$ m.u.) density field 
analyzed is the largest cubic subregion of the original output.}. If we 
consider the percolation threshold to be the first significant jump in the 
value of the largest structure statistic (\cite{deL-Gel-Huc86}) then 
percolation happens for all versions in the range $0.024 \le f\!f \le 
0.05$ for clusters and between $0.01 \le f\!f \le 0.022$ for voids. For cluster 
percolation,  fields with larger values of $\beta$ percolate at smaller 
filling factor values, while for voids the trend is generally reversed. 
This means the larger the average cosmic density or the smaller the bias 
factor the more connected the clusters tend to be.  The fact that the 
percolation threshold is the distinguishing difference between the curves 
demonstrates the sensitivity of this parameter as suggested by Shandarin 
(1983).  Alternately, the high sensitivity may also cause problems in noisy 
samples as reported by Dekel \& West (1985).  However,
rigorously determining a percolation threshold value is not important to 
the analysis in this study and is used here only for illustration. In this 
study the shape of the largest structure function over its entire range will 
be used as a comparison to characterize the topology of a field. 

Another important feature of the largest structure statistic is the high 
initial values at low filling factors for all versions of the Wiener 
Reconstructions. This indicates that the largest structure 
is always dominant which indicates a problem with the size of the 
sample. In a statistically fair sample, there would be a multitude of small 
clusters at the high 
density cutoffs beginning the percolation process, and the largest 
cluster would emerge from the field as the percolation process caused clusters 
to join together.  The fact that the statistic has a non-vanishing initial 
value is indicative of the relatively small sample size of this survey for the 
purpose of this statistic and the level of smoothing introduced by the 
reconstruction process. For comparison, see the percolation results of N-body 
simulations in Figure 3 and reference Yess \& Shandarin (1996).  

A lack of resolution at large distances explains the relationship between the 
results of the upper and lower panels of Figure 1.  The growth of the largest 
structure function is virtually identical for the two cases except for a near 
doubling in the 
filling factor. This means that all information about structure is contained 
in the reduced spherical region bounded by $R\approx 100 h^{-1} $ Mpc 
(30 mu), and that 
the excess volume in the cube does not affect the structures but only 
contributes to a reduction in the filling factor. The implication is that the 
value of the field outside $R\approx 100 h^{-1} $ Mpc is featureless and 
roughly equal to the mean density. This is because attenuation of the signal 
is a function of radius due to the effective smoothing of the spherical 
harmonic reconstruction and the loss of detail after Wiener filtering. The 
effects of these two operations will be examined separately below. 

Restricting the analysis to an even smaller volume ($R\approx 30 h^{-1} $ Mpc  
(10 mu), 
not shown), reveals similar percolation curves to those of larger volumes for 
the growth of the largest clusters except that volume effects for clusters 
are exaggerated to the point where the largest cluster is 
associated with the highest density peak and its volume is never less than 
half the volume of all clusters combined. For voids the curves are also 
similar at the outset, but rise much slower so $f\!f \approx 0.3$ when the 
volume of the largest void approaches unity.

In Figure 2, we display the results of a systematic study of the effects of 
spherical harmonic reconstruction and Wiener filtering separately on N-body 
simulations. The results of analysis of four realizations of N-body 
simulations characterized by an initial power law power spectrum of the form 
$P(k) \propto k^{-1}$, evolved to the stage where $\lambda = \lambda_f/8$  
(where $\lambda_f$ is the fundamental wavelength) is approaching non-linearity 
\footnote{ For a detailed discussion of the N-body simulations used in this 
study see Melott \& Shandarin (1993).}. 
The upper panels show the results of percolation analysis for the 
simulations and demonstrate that the topology of the structure is similar 
throughout the volume and not a function of radius.  The difference between 
percolation in a cube and a sphere is also insignificant.  It is also 
important to note that the rapid growth of the largest structure for random 
Gaussian fields (light lines) starts at a filling factor of 
$f\!f \approx 0.16$ for both clusters and voids, which is the expected value.  
The interpretation of this data is that the topology of these 
simulations is characterized by a very connected cluster network and 
slightly more isolated voids compared to Gaussian fields.  

The middle panels show the effect of spherical harmonic reconstruction alone 
and in conjunction with Wiener filtering. The reconstruction process 
eliminates the distinction between the topologies of clusters, voids and  the 
random fields in a spherical volume of radius, $R \approx 100 h^{-1} $ Mpc, 
and introduces some distortion in the curves at low filling factors. 
In addition, the Wiener filter removes most of the small scale structure 
demonstrated by the almost immediate percolation of the fields and the 
reduction of error bars. It is easy to understand this effect of the Wiener 
filter because it reduces the range of the density values in the 
reconstructed density field by one-fourth. In order to regain a measure 
similar to that of the original density fields, the effects of the 
reconstruction and filtering have to be minimized by restricting the analysis 
to smaller radii.  The smoothing effect of the reconstruction is less and the 
signal to noise ratio is better for smaller radii, so that the topology of the 
original field can be recognized in a volume limited sample at 
$R \approx 30h^{-1} $ Mpc if 
the reconstruction alone is applied to the simulation (bottom left panel). The 
Wiener filtering still distorts the original topology even in this restricted 
volume as shown in the bottom right panel. An important feature for the 
interpretation of percolation results is that the largest structures in the 
random Gaussian realizations behave, making allowance for survey volume 
effects and smoothing, generally as expected in all cases except the middle 
right panel. 

Finally, Figure 3 shows the largest structure statistics for Wiener 
reconstructions of the IRAS 1.2 Jy Survey for various $\beta$ values. 
Each version displays a similar result with voids percolating at lower filling 
factors than the Gaussian counterparts, and clusters percolating similarly to 
the Gaussian fields. These results imply a well connected void distribution 
with a generally sponge like or slightly meatball cluster distribution. 
Volume and smoothing effects are again evident in each field 
demonstrated by the high values of the largest structure statistic at low 
filling factor values. This problem is more apparent in the cluster 
analysis of the original IRAS reconstructions than for voids or clusters in 
randomized IRAS reconstructions. Although the lack of resolution prevents a 
strong characterization of the topologies represented in the data, the results 
are consistent with the slight meatball topology shift reported by Moore 
{\it et al.} (1992) and Gott {\it et al.} (1989) for similar local volumes. 
Results for volumes with $R \approx 30 h^{-1} $ Mpc (not shown) are 
inconclusive.

\subsection{Number of Clusters}

It has been established that the maximum of the number of clusters statistic 
reflects the slope of the power spectrum for density fields described by a 
simple power law (\cite{Yes-Sh96}). This is true for Gaussian fields and 
randomized density fields derived from N-body simulations over a wide range 
of evolutionary stages. This statistic can easily be normalized by the volume 
of the field so that different samples can be directly compared; however, the 
values of the maximum are small enough in this study that the raw data is 
presented for clarity. The results 
of percolation analysis of clusters and voids are presented in Figure 4 for
both reconstructed IRAS survey ($\beta= 0.1, 0.5, 1.0$) and N-body simulation 
($n= 0, -1, -2$) fields. The most notable 
feature of the data is that the maxima for N-body simulation fields are two 
orders of magnitude below those of the original fields before they were 
reconstructed. 
This reduction in the number of clusters (voids) is a direct result of 
the smoothing and attenuation of the Wiener Reconstruction procedure. The 
effect is to reduce the signal below the level where distinctions can be made 
between fields characterized by different spectra. Another indication that the 
resolution of the reconstructions is insufficient for percolation analysis are 
the many local maxima in the statistic.

\section{Conclusions}

The Wiener reconstruction technique has 
proven successful in reconstructing the angular density fields of galaxies 
(\cite{Lah-etal94}); the temperature fluctuations of the Cosmic Microwave 
Background (\cite{Bun-etal94}); real space density, velocity and gravitational 
potential fields (\cite{Fis-etal95a}); and predicted full sky density fields 
(\cite{Zar-etal95}).  In this study we apply percolation analysis to full sky 
Wiener Reconstructions of the IRAS 1.2 Jy Redshift Survey. We find that 
our results are consistent with the conclusions of other studies that report 
a small shift towards a meatball topology for the survey region,
however we would like to stress that our analysis was in real space.

The Wiener reconstruction technique smoothes the density field based on the 
inter-galaxy separation as a function of radius and attenuates the high 
frequency components or small scale components resulting from shot noise. 
This results in a loss of resolution with distance which presents 
a challenge for percolation analysis.  The largest structure statistic is 
robust enough to give an indication of the topology of the field under these 
conditions; however, the number of clusters statistic suffers too much from 
the loss of resolution to give a measure of the slope of the associated 
power spectrum. Alternately, the number of clusters statistic has the potential 
to be developed into an indicator of whether or not the structures of a given 
field represent a fair sample for statistical purposes. 

The prospects of analyzing a full sky density reconstruction in order to 
assess the topology of the large scale structure and associate that structure  
with initial fluctuations in the matter density field at the time of 
recombination are attractive. This study offers an optimistic picture that as 
more galaxies are added to surveys the statistical measures presented will 
produce accurate and convincing results.  Until galaxy survey counts are 
increased enough to overcome the problems identified in this study, 
percolation can still be applied to point-wise galaxy distributions, density 
fields of highly sampled portions of the sky or ideally volume limited 
subsamples.  

S. Shandarin acknowledges support from NASA Grant NAGW-3832, NSF Grant 
AST-9021414 and the University of Kansas Grant GRF96.  We would like to thank 
David Weinberg for valuable suggestions and comments.

\vfill

\newpage

\begin{thebibliography}{}

\baselineskip 0.8cm

\bibitem[Baugh 1993]{Bau93}
Baugh, C. 1993, \mnras {264} {87}
\bibitem[Bhavsar \& Splinter 1996]{Bha-Spl96}
Bhavsar, S., \& Splinter R., (1996) in progress
\bibitem[Bunn {\it et al}. 1994]{Bun-etal94}
Bunn, E., Fisher, K. B., Hoffman, Y., Lahav, O., Silk, J., \& Zaroubi, S.
1994, \apjl {432} {L75}
\bibitem[Davis {\it et al}. 1982]{Dav-etal82} 
Davis, M., Huchra, J., Latham, D. W. \& Tonry, J. 1982. \apj {253} {423}
\bibitem[Dekel \& West 1985]{Dek-Wes85} 
Dekel, A., \& West, M. J. 1985, \apj {288} {411}
\bibitem[Fisher {\it et al}. 1995a]{Fis-etal95a}
Fisher, K.B., Lahav, O., Hoffman, Y., Lynden-Bell, D. \& Zaroubi, S. 1995a
\mnras {272} {885}
\bibitem[Fisher {\it et al}. 1995b]{Fis-etal95b}
Fisher, K. B., Huchra J. P., Strauss, M. A., Davis M., Yahil, A., \& 
Schlegel, D. 1995b, \apjs {100} {69}
\bibitem[Geller \& Huchra 1989]{Gel-Huc89}
Geller, M. J. \& Huchra, J. 1989, {\it Science}, 246, 897.
\bibitem[Giovanelli \& Haynes 1985]{Gio-Hay85}
Giovanelli, R., \& Haynes, M. P. 1985, \aj {90} {2445}
\bibitem[Gott, Melott \& Dickinson 1986]{Got-Mel-Dic86} 
Gott, J. R., Melott, A. L., \& Dickinson, M. 1986, \apj {306} {341}
\bibitem[Gott {\it et al}. 1989]{Got-etal89}
Gott, J. R., Miller, J., Thuan, T. X., Schneider, S. E., Weinberg, D. H., 
Gammie, C., Polk, K., Vogeley, M., Jeffery, S., Bhavsar, S., Melott, A. L., 
Giovanelli, R., Haynes, M. P., Tully, R. \& Hamiltion, A. 1989, 
\apj {340} {625}
\bibitem[Gunn \& Weinberg 1995]{Gun-Wei95}
Gunn, J. E. \& Weinberg, D. H. 1995, {\it Wide-Field Spectroscopy and the 
Distant Universe, eds}. S. J. Maddox and A. $Arag\acute{o}n-Salamanca$, 
(Singapore : World Scientific).
\bibitem[Kirshner {\it et al}. 1981]{Kir-etal81}
Kirshner, R. P., Oemler, A., Schechter, P. L., \& Shectman, S. A. 1981, 
\apj {248} {L57}
\bibitem[Lahav {\it et al}. 1994]{Lah-etal94}
Lahav, O., Fisher, K. B., Hoffman, Y., Schaarf, C. A., \& Zaroubi, S. 1994
\apjl {423} {L93}
\bibitem[de Lapparent, Geller \& Huchra 1986]{deL-Gel-Huc86} 
de Lapparent, V., Geller, M., \& Huchra, J. P. 1986, \apj {301} {L1}
\bibitem[de Lapparent, Geller \& Huchra 1991]{deL-Gel-Huc91} 
de Lapparent, V., Geller, M., \& Huchra, J. P. 1991, \apj {369} {273}
\bibitem[Lawrance {\it et al}. 1996]{Law-etal96}
Lawrence, A., Rowan-Robinson, M., Crawford, J., Saunders, W., Ellis, R. S., 
Frenk, C., Parry, I., Efstathiou, G., \& Kaiser, N. 1995 in preparation
\bibitem[Mecke, Buchert \& Wagner 1994]{Mec-Buc-Wag94}
Mecke, K., Buchert, T. \& Wagner, H. 1994, \aa {288} {697}
\bibitem[Melott \& Shandarin 1993]{Mel-Sh93} 
Melott, A. L., \& Shandarin, S. F.  1993, \apj {410} {469}
\bibitem[Moore {\it et al}. 1992]{Moo-etal92} 
Moore, B., Frenk, C. S., Weinberg, D. H., Saunders, W., Lawrence, A., 
Ellis, R. S., Kaiser, N., Efstathiou, G., \& Rowan-Robinson, M.  1992, 
\mnras {256} {477}
\bibitem[Peebles 1980]{Pee80}
Peebles, P. J. E., 1980. {\it The Large Scale Structure of the Universe}, 
Princeton University Press, Princeton, NJ.
\bibitem[Ryden 1988]{Ryd88}
Ryden, B. S. 1988, \apj {333} {L41}
\bibitem[Rybicki \& Press 1992]{Ryb-Pre92}
Rybicki, G. B., \& Press, W. H. 1992, \apj {398} {169}
\bibitem[Sahni, Sathyaprakash, \& Shandarin 1994]{Sah-Sat-Sh94}
Sahni, V., Sathyaprakash, B. S., \& Shandarin, S. F. 1994, \apj {431} {20}
\bibitem[Sathyaprakash, Sahni, \& Shandarin 1995]{Sat-Sah-Sh96} 
Sathyaprakash, B. S., Sahni, V., \& Shandarin, S. F. 1996, \apjl {462} {L5}
\bibitem[Shandarin 1983]{Sh83} 
Shandarin, S. F. 1983, \sastl {9} {104}
\bibitem[Shandarin \& Zel'dovich 1983]{Sh-Z83} 
Shandarin, S. F., \& Zel'dovich, Ya. B. 1983, \com {10} {33}
\bibitem[Shane \& Wirtanen 1967]{Sha-Wir67}
Shane, C. D. \& Wirtanen, C. A., 1967. {\it Publ.Lick Obs.,} 22, 1.
\bibitem[Strauss {\it et al}. 1992]{Str-etal92}
Strauss, M. A., Huchra, J. P., Davis, M., Yahil, A., Fisher, K., \& Tonry, J. 
1992a, \apjs {83} {29}
\bibitem[Tully \& Fisher 1987]{Tul-Fis87}
Tully, R. B., \& Fisher, J. R. 1987, {\it Nearby Galaxies Atlas} (Cambridge: 
Cambridge University Press).
\bibitem[Vogeley {\it et al}. 1994]{Vog-etal94} 
Vogeley, M. S., Park, C., Geller, M. J., Huchra, J. P., \& Gott, J.R. 1994, 
\apj {420} {525}
\bibitem[Yess \& Shandarin 1996]{Yes-Sh96}
Yess, C., \& Shandarin S. F. 1996, \apj {465} July 1 issue
\bibitem[Zaroubi {\it et al}. 1995]{Zar-etal95}
Zaroubi, S., Hoffmean, Y., Fisher, K. B., \& Lahav, O. 1995, \apj {449} {446}
\bibitem[Zel'dovich 1982]{Z82} 
Zel'dovich, Ya. B. 1982, \sastl {8} {102}
\bibitem[Zwicky {\it et al}. 1961-68]{Zwi61-68}
Zwicky, F., Herzog, E., Wild, P., Karpowicz, M. \& Kowai, C. T., 1961-1968. 
{\it Catalogue of Galaxies and Clusters of Galaxies}, California Institute of 
Technology, Pasadena



\end{thebibliography}
\end{document}